\UseRawInputEncoding
\documentclass[prb,twocolumn,showpacs,preprintnumbers,amsmath,amssymb]{revtex4}
\usepackage{graphicx}
\usepackage{dcolumn}
\usepackage{bm}
\usepackage[tight]{subfigure}
\usepackage{enumerate}
\usepackage{amsmath}
\usepackage{verbatim}
\usepackage{color}
\usepackage{lipsum}

\begin{document}

\title{Thermoelectric properties of armchair phosphorene nanoribbons in the presence of vacancy-induced impurity band}
\author{Mohsen Rezaei}
\author{Hossein Karbaschi}
\email{h.karbaschi@gmail.com}
\author{Mohsen Amini}
\email{msn.amini@sci.ui.ac.ir}
\author{Morteza Soltani}
\author{Gholamreza Rashedi}

\affiliation{Department of Physics, University of Isfahan, Isfahan 81746-73441, Iran} 

\begin{abstract}
Armchair phosphorene nanoribbons (APNRs) are known to be semiconductors with an indirect bandgap. Here, we propose to introduce new states in the gap of APNRs by creating a periodic
structure of vacancies (antidots). Based on the tight-binding model, we show that a periodic array
of vacancies or nanopores leads to the formation of an impurity band inside the gap region. We first present an analytical expression for the dispersion relation of an impurity band induced by hybridization of bound states associated with each single vacancy defect. Then, we increase the size of vacancy defects to include a bunch of atoms and theoretically investigate the effect of nanopores size and their spacing on electronic band structure, carrier transmission function, and thermoelectric
properties. Our analysis of the power generation rate and thermoelectric efficiency of these structures reveals that an ANPR can be used as a superb thermoelectric power generation module.
\end{abstract}
\maketitle

\section{Introduction}

Thermoelectric (TE) modules refer to classes of solid-state power generators that convert heat current directly into electrical current or thermoelectric cooling (heating) devices which create a temperature gradient by consuming electrical power. 
Therefore, TE materials have an essential role in the development of both 
electrical power generators and heating (cooling) modules.
The TE power generation occurs in semiconductors and metals due to their charge carrier’s mobility. 
In such materials, an applied temperature gradient  
will cause the charge carriers to diffuse 
from hot to cold side, producing an electrical current. 
The most important advantage of thermoelectric generators (TEG) over traditional power generators is the ability to derive their power from waste heat even at small temperature gradients. 
Furthermore, TEGs contain no moving parts, making them reliable and maintenance-free. 
They are also silent, emission-free, and flexible in architecture.

Although the study of the power generation properties of TE materials is an old subject, and despite the many advantages of TEGs over conventional systems, their usage is presently limited to some special cases, due to their low efficiencies.

To improve the efficiency of TEGs, researchers have proposed a variety of methods. The idea of using Nano-structured materials is one of the most successful solutions to enhance the efficiency of TEG~\cite{Hicks93a, Hicks93b}. The enhancement in the performance of TEGs in low dimensional systems is mainly originated both from quantum effects relevant to the confinement of carriers and also from the considerable scattering of phonons at the boundaries. There have already been a lot of experimental and theoretical studies focused on improving the thermoelectric power generation efficiency in low dimensional structures~\cite{Esposito09,Whitney14, Whitney15,Wu2013,Miyazaki2013,Zhao2014,Liu2014,JinBae2016,ZHANG2016,Murphy08,Leijnse10,Karbaschi2016,Hung2016,Karbaschi2020}. 


Recently, two-dimensional (2D) materials have attracted lots of attention due to their distinguished properties which considerably differ from corresponding 3D materials. 
Among the newly developed 2D materials, phosphorene stands out due to its high carrier mobility, a large on: off ratio, and a tunable band gap~\cite{Li2014, Liu2014_2, Long2016}.
Furthermore, it is reported that at room temperature, the values of the Seebeck coefficient of bulk black phosphorus and the figure of merit of phosphorene reaches $345~\mu V/K$ and $1.0$, respectively~\cite{Flores2015,Fei2014}.
Recently, both theoretical and experimental studies have investigated the possibility of TE efficiency enhancement in a strip of phosphorene which is called phosphorene nanoribbon~\cite{Zhang2014}.
Additionally, it is shown that the lattice thermal conductivity of a phosphorene nanoribbon is much smaller than that of graphene~\cite{Hong2015}. It is also predicted that the thermal conductivity of APNR is 2-3 times smaller than that of zigzag phosphorene nanoribbon (ZPNR)~\cite{ZHANG2016, Qin2015}.
On the other hand, defect engineering has also emerged as a mechanism to enhance the thermoelectric properties of different materials extending the scope of their application beyond the pristine structure~\cite{Chang2014}.
Promising results are reported with thin films~\cite{Tang2010}, graphene and finite graphene antidot lattices~\cite{Xu2019, Gunst2011}, and multi-layer graphene nanomeshes~\cite{Oh2017}. 
It is, therefore, crucial to study the effects of the periodic array of nanoholes embedded into APNRs on the thermoelectric properties to achieve the desired performance.
 
Taking this as motivation, in this work, a further enhancement of thermoelectric performance is proposed by creating superlattice defects and engineering nanopores in APNRs which is shown to be experimentally possible in phosphorene systems recently~\cite{Cupo2017}. 
We theoretically study the effects of superlattice defects, namely, vacancies and nanopores defects on the thermoelectric properties of APNRs.
We show that the introduction of a line of vacancy defects can induce mid-gap impurity states in APNRs and, thus, a new impurity band may be formed in the gap region. 
To understand the mechanism of impurity band formation, we employ the tight-binding model for a periodic chain-like structure of single vacancy defects and obtain an analytical expression for the dispersion relation of the emergent impurity band.
Using this knowledge, we further increase the size of the vacancy defects to nanopores and calculate the carrier's band structure and transmission function.
Furthermore, we discuss the effect of the nanoholes' size and their spacing on the transmission function, electrical power output, and thermoelectric efficiency.
We find that, depending on the size of patterns and the repeat periodicity of vacancies, it is possible to reach high electrical conductivity. 
Furthermore, previous studies have shown that the presence of defects in the crystal structure of nanoribbons can dramatically reduce the thermal conductivity of the system~\cite{SharafatHossain2015,Karamitaheri2012,YAN2012}, which results in an effective enhancement of the thermoelectric efficiency of the system.

The rest of the paper is organized as follows: In section~\ref{sec:model} we present our defective model and method which 
we use in this study. We also discuss how an impurity band form in the gap region of the APNR when an array of periodic vacancy defects is considered. We then generalized our study to the case of larger vacancy defects and discuss the  main TE properties of the system
using the exact calculations based on the Landauer-B$\mathrm{\ddot{u}}$ttiker formula in section~ \ref{sec:results}. Finally, we summarize our work and end with some conclusions in Section~\ref{sec:conclusions}.

\begin{figure}[]
  	\includegraphics[height=1.0\linewidth]{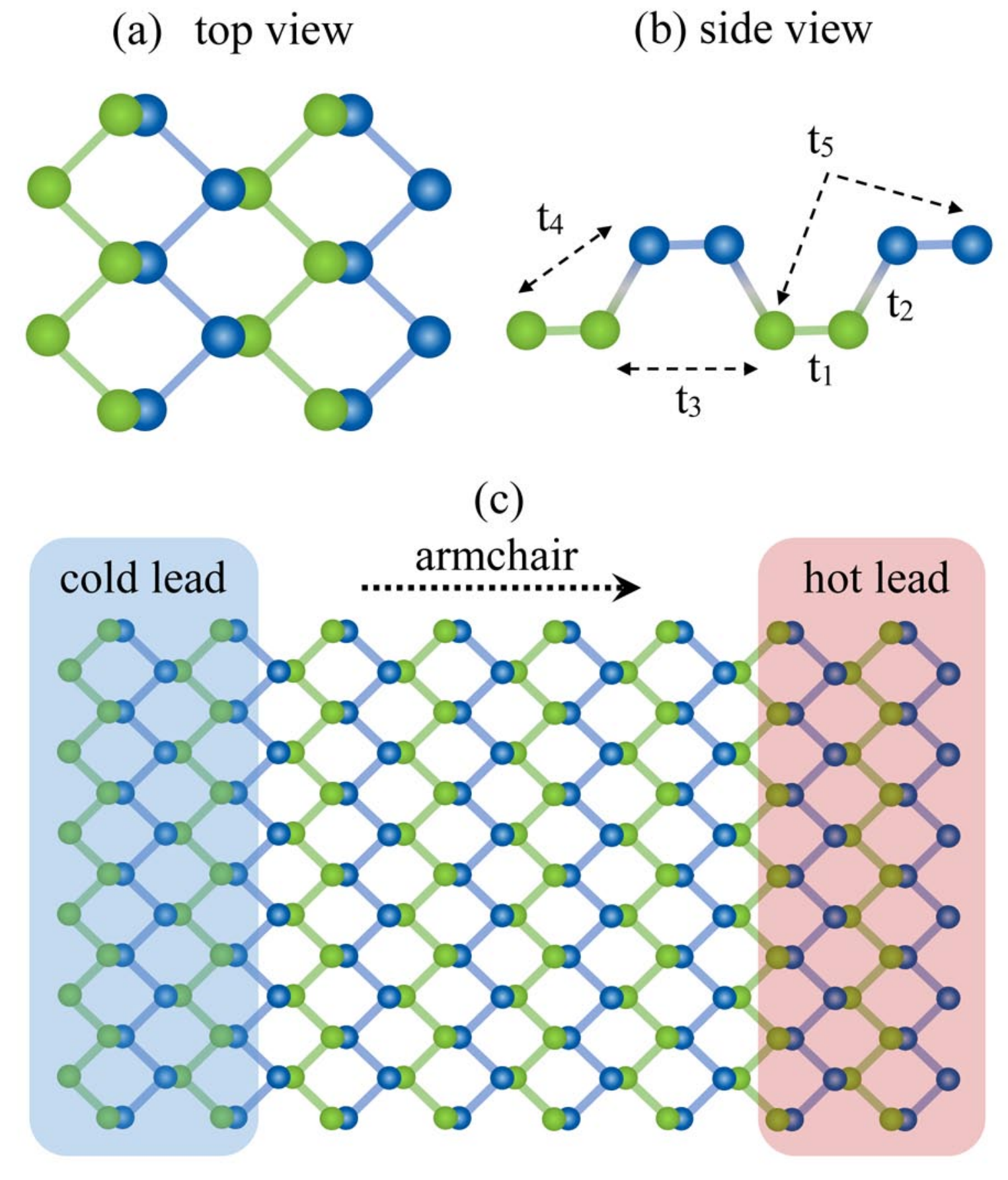}	
	\caption{\label{fig:setup} 
	(a) Top and (b)  side view of the phosphorene crystal structure, (c) armchair phosphorene nanoribbon as a scattering region attached to hot and cold leads.
	 }
\end{figure}

\section{Model and methods \label{sec:model}}
\subsection{Preliminary concepts}
The crystal structure of a monolayer phosphorene is presented in Figures~\ref{fig:setup} (a) and (b) which illustrate the top and side view of its nonplanar puckered honeycomb lattice, respectively.
Each lattice site belongs to either $A$ or $B$ sublattices and can be labeled with a set $(n, m,\nu)$, where
$n$ and  $m$ represent the $x$ and $y$ index of the lattice points and $\nu=A,B$ refers to the sublattice  (see Fig.~\ref{fig:setup} (a)).
We use the tight-binding Hamiltonian, which accurately describes the phosphorene structures~\cite{Ezawa2014,Rudenko2014}
and can be written as
\begin{equation}
{\hat H} =  \sum_{ i \neq j } t_{ij} c_{i}^\dagger c_j + H.c. , \label{EQ1}
\end{equation}
where the second-quantized operator $c^\dagger_i(c_i)$ is the electron creation (annihilation)  operator in site $i$, $t_{ij}$ represents the energy cost for an electron to hop between sites $i$ and $j$, and $H.c.$ stands for hermitian conjugate.
While it is sufficient~\cite{Rudenko2014,Rudenko2015} to describe the low-energy region of the band structure of phosphorene by considering the hopping integrals up to the fifth-nearest neighbors (see Fig.~\ref{fig:setup} (b)), 
we only keep the important hopping terms to the
first-, second-, and fourth-nearest neighbors in our analytic calculations~\cite{Ezawa2014, Amini2019_1, Amini2019_2}.
The values of these hopping integrals are
$t_1 =-1.220$~eV, $t_2 = 3.665$~eV, $t_3 =-0.205$~eV, $t_4 =-0.105$~eV, and $t_5 =-0.055$~eV~\cite{Rudenko2014}.
Therefore, we can write the Hamiltonian as
\begin{equation}
\label{EQ2}
\begin{split}
{\hat H} &  = {\hat H_0}+{\hat H_1}, \\
{\hat H_0} &  = t_1 \sum_{\langle i,j \rangle_\text{1st}}  c_{i}^\dagger c_j + t_2 \sum_{\langle i,j \rangle_\text{2nd}}  c_{i}^\dagger c_j + H.c., \\
{\hat H_1}  &  = t_4 \sum_{\langle i,j \rangle_\text{4th}}  c_{i}^\dagger c_j + H.c.
\end{split}
\end{equation}

The structure of the APNR can be considered as a stripe of phosphorene with infinite length in the $x$ direction and finite width in the $y$ direction which is presented in Fig.~\ref{fig:setup} (c).
The armchair edge of phosphorene does not support edge states, therefore pristine APNRs are semiconductors with an indirect band gap~\cite{Ezawa2014}.
In our thermoelectric studies, we consider an APNR as a scattering region attached to two hot and cold leads, as shown in Fig.~\ref{fig:setup} (c). The hot and cold leads are considered as electron Fermi seas with temperatures ($T_h$) and ($T_c$) and electrochemical potentials $\mu_h$ and $\mu_c$, respectively.

\subsection{Impurity bands in armchair phosphorene antidot nanoribbons}

We now consider a vacancy superlattice of APNR by creating a periodic array of vacancies with a given separation between the centers of two consecutive holes to form chain-like line defects.
However, before considering defective APNR, it is insightful to overview the effect of a single vacancy defect in bulk phosphorene first.  
A single vacancy defect can be described by switching off all the hoppings between the vacant site and its neighbors. 
It is known that a single vacancy defect in the bulk phosphorene induces an impurity state near the
vacancy position which shows a highly anisotropic localization in real space~\cite{Amini2019_2,Kiraly2017}.
Following the notation of Refs.~\cite{Amini2019_2, Amini2021}, if such vacancy defect is located on a site that belongs to the $B$ sublattice, the associated wavefunction with this impurity will be shown as $\mid\psi^{A}\rangle$.

\begin{figure}[]
  	\includegraphics[height=0.68\linewidth]{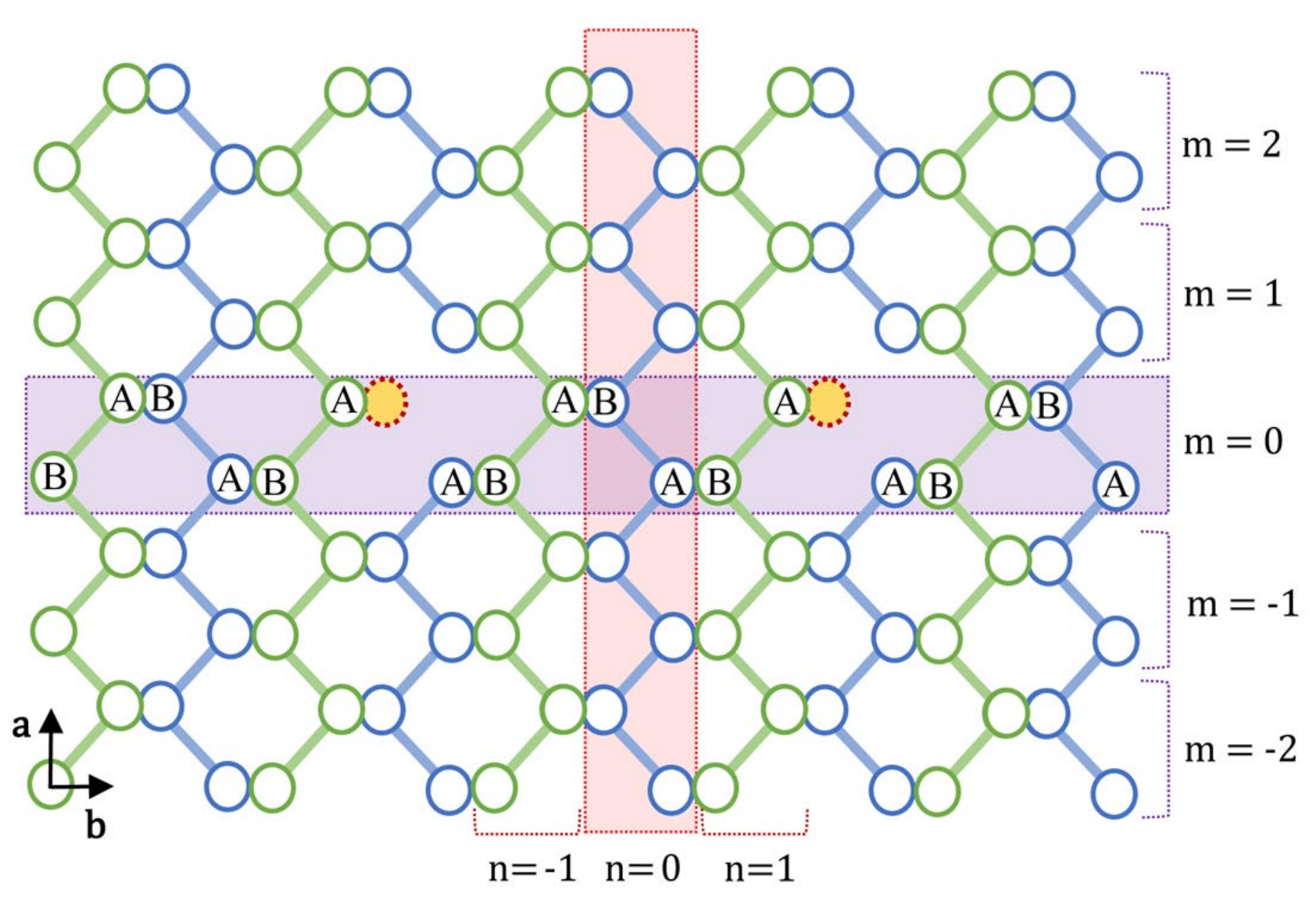}	
	\caption{\label{fig:supercell} 
	Schematic top view of armchair phosphorene nanoribbon. Here, $n$ and $m$ denote the armchair chain  and supercell numbers, respectively.
	}
\end{figure}

Let us,  now, consider an array of single vacancies introduced in a periodic chain structure at the middle of the APNR. 
In this way, a vacancy lattice is defined as a line of $N$ vacancy defects which are located on sites $(jn_0,0,B)$ for $j=0,1,...,N-1$.
The intervacancy distance is therefore $n_0b$ where $b$ is the phosphorene lattice spacing along the horizontal axis
(the case with $n_0=4$ is schematically represented in Fig.~\ref{fig:supercell}).
If the intervacancy distance as well as the width of the ribbon is sufficiently large (in comparison to the decay length of the impurity state), we can still describe each single site on the vacancy lattice using the vacancy wavefunction $\mid\psi^{A}\rangle$.
Using the notation of Ref.~\cite{Amini2019_2}, it can be written as
\begin{equation}
\label{EQ3}
|\psi_j^{A}\rangle = c \int_{-\pi}^{\pi} \gamma^{-1}({k_y})|\Psi_j^A(k_y) \rangle dk_y,
\end{equation}
where $c$ is the normalization prefactor given by
\begin{equation}
\label{EQ4}
c^{-2} = \int_{-\pi}^{\pi} \gamma^{-2}(k_y) dk_y,
\end{equation}  
and
\begin{equation}
|\Psi_j^A(k_y)\rangle=\frac{1}{\sqrt{\pi}}\sum_{n,m}^\prime  \alpha^n(k_y) \gamma(k_y)e^{ik_y(m-\delta_n)}|n-n_0j,m,A\rangle.
\label{EQ5}
\end{equation}
The prime indicates that the summation over $n$ is only restricted to sites which are located on the left side of the $j$-th vacancy,
$\delta_n$ is a constant parameter equal to 0~(0.5) for even~(odd) $n$, 
and
\begin{equation}
\label{EQ6}
\gamma(k_y)=\sqrt{1-\alpha^2(k_y)}=\sqrt{1-(2(t_1/t_2)\cos{(k_y/2))^2}}.
\end{equation} 
In these expressions, the wave-vector $k_y$ is measured in units of the inverse lattice spacing $a$ along the vertical axis (see Fig.~\ref{fig:supercell}).

It is obvious that the above-mentioned vacancy bound states can hybridize through
the kinetic hopping term $H_1$ in Eq.~(\ref{EQ2}) to form an impurity band.
In order to obtain the dispersion relation of such impurity band we construct the following wavefunctions
\begin{equation}
\label{EQ7}
 \mid\phi(k_n)\rangle= \frac{1}{\beta(k_n)}\sum_je^{ik_nj}\mid\psi_j^A\rangle,
\end{equation} 
which is equivalent to the Fourier expansion of wavefunctions $\mid\psi_j^A\rangle$ when $N\rightarrow\infty$.
Here, $k_n=\frac{2\pi n}{N}, n=0,1,...,N-1$ and $\beta(k_n)$  represents the overall normalization factor which fulfils the following
requirement
\begin{equation}
\label{EQ8}
\langle\phi(k_n)\mid\phi(k_n)\rangle=\frac{1}{\beta^2(k_n)}\sum_{j,j^\prime}e^{ik_n(j-j^\prime)}\langle\psi^A_{j^\prime}\mid\psi^A_{j}\rangle=1,
\end{equation}
and results in
\begin{equation}
\label{EQ9}
 \beta^2(k_n)=\sum_{j,j^\prime}e^{ik_n(j-j^\prime)}\langle\psi^A_{j^\prime}\mid\psi^A_{j}\rangle.
\end{equation}

It is worth noting that the wavefunctions of Eq.~(\ref{EQ3}) are not orthogonal to each other, namely $\langle\psi_{i}^A\mid\psi_{j}^A\rangle\neq 0$,  while the one defined in Eq.~(\ref{EQ7}) obey the orthogonality relations as $\langle\phi(k_i)\mid \phi(k_j)\rangle=\delta_{ij}$
(the details of the orthogonality properties are presented in Appendix~\ref{APA}).
The energy dispersion relation of the impurity band, $E(k)$, is therefore given by 
\begin{equation}
\label{EQ10}
\begin{split}
E(k_n) &=\langle\phi(k_n)\mid H_1\mid\phi(k_n)\rangle \\
 &=\frac{1}{\beta^2(k_n)}\sum_{j,j^\prime}e^{ik_n(j-j^\prime)}\langle\psi^A_{j^\prime}\mid H_1\mid\psi^A_{j}\rangle \\
 &=\frac{\sum_{j,j^\prime}e^{ik_n(j-j^\prime)}\langle\psi^A_{j^\prime}\mid H_1\mid\psi^A_{j}\rangle}{\sum_{j,j^\prime}e^{ik_n(j-j^\prime)}\langle\psi^A_{j^\prime}\mid\psi^A_{j}\rangle}.
\end{split}
\end{equation}
For the sake of simplicity, let us assume $j^\prime=0$. Then, $E(k_n)$ take the form
\begin{equation}
\label{EQ11}
E(k_n) =\frac{\langle\psi^A_{0}\mid H_1\mid\psi^A_{0}\rangle+2\sum_{j}\langle\psi^A_{0}\mid H_1\mid\psi^A_{j}\rangle\cos(k_nj)}{1+2\sum_{j}\langle\psi^A_{0}\mid\psi^A_{j}\rangle\cos(k_nj)},
\end{equation}
in which the expectation values in the numerator can be evaluated using Eq.~(\ref{EQ3}) as
\begin{align} 
\label{EQ12}
\langle&\psi^A_{0}\mid H_1\mid\psi^A_{0}\rangle=\nonumber\\&c^{-2}\int^\pi_{-\pi}\gamma^{-1}(k_y)\gamma^{-1}(k^\prime_y)\langle\Psi^A_{0}(k^\prime_y)\mid H_1\mid\Psi^A_{0}(k_y)\rangle dk_ydk_y^\prime\nonumber\\&=c^{-2}\int^\pi_{-\pi}\gamma^{-2}(k_y)(-4t^\prime) \cos^2(k_y/2)dk_y,
\end{align} 
and
\begin{align}
\label{EQ13}
\langle&\psi^A_{0}\mid H_1\mid\psi^A_{j}\rangle=\nonumber\\&c^{-2}\int^\pi_{-\pi}\gamma^{-2}(k_y)\alpha^{n_0j}(k_y)(-4t^\prime) \cos^2(k_y/2)dk_y,
\end{align} 
with $t^\prime=2t_1t_4/t_2$.
Similarly, the overlap integral in the denominator reads
\begin{align}
\label{EQ14}
\langle\psi^A_{0}\mid\psi^A_{j}\rangle=c^{-2}\int^\pi_{-\pi}\gamma^{-2}(k_y)\alpha^{n_0j}(k_y)dk_y.
\end{align}

\begin{figure}[]
  	\includegraphics[height=1.2\linewidth]{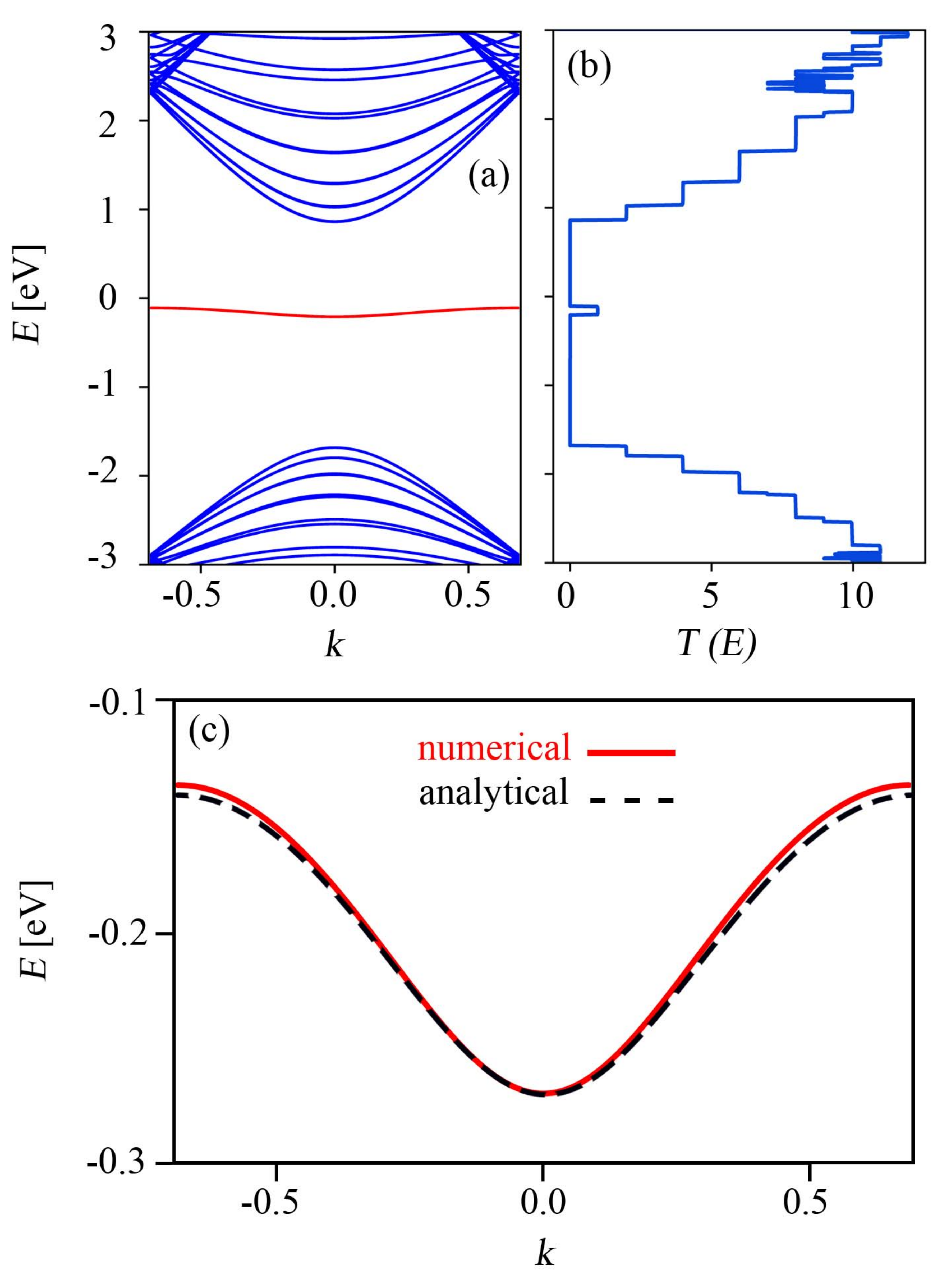}	
	\caption{\label{fig:bandstructure} 
	(a) Bandstructure and (b) transmission function of armchair phosphoren nanoribbon with periodic vacancy. (c) comparison between the analytical and numerical calculations of impurity band.
 }
\end{figure}

Finally, we can easily perform the integrations of Eqs.~(\ref{EQ12}),~(\ref{EQ13}),~(\ref{EQ14}) to obtain the dispersion relation of the impurity band in Eq.~(\ref{EQ11}).
A very good agreement is obtained between the obtained dispersion relation of Eq.~(\ref{EQ11}) and the calculated band structure in numerical calculations for the case of $n_0=4 (d=4b)$ which is represented in 
Fig.~(\ref{fig:bandstructure}) (c).



\subsection{Thermoelectric efficiency \label{sec:eff}}

The efficiency of the thermoelectric power generators is defined as the ratio of the generated electrical power (output) to the rate of heat which is lost from the hot lead (input). Economically point of view, the best TEGs are those that generate the highest electrical power with the least heat exchange.
\begin{equation}\label{eq:cop}
	\eta = \frac{P_{out}}{P_{in}}=\frac{P}{\dot{Q}_h}=\frac{VI}{\dot{Q}_h},
\end{equation}

Where $V$ denotes the bias voltage. Using the Landauer-B$\mathrm{\ddot{u}}$ttiker formalism, one can calculate the electric current. In the ballistic transport regime, the expression for calculating the current will then be
\begin{equation}\label{eq:current}
	I = \frac{2 e}{h} \int dE \; T_{LR}(E) \left( f_L-f_R\right),
\end{equation}

Here, $h$ and $e$ are the Planck constant and the elemental charge of one electron, respectively. $T(E)$ is the total transmission function, which in this study, have been calculated using the KWANT package \cite{Groth2014}. Also,   $f=1/\left[ e^{(E-E_F)/k_BT}+1\right]$ is the Fermi-Dirac distribution function and $k_B$ is Boltzmann constant.

The experission for calculating the electron contribution of heat current is the same as electric current but replacing the elemental charge $e$ with the $(E-\mu_h)$. 

\begin{equation}\label{eq:heatcurrent}
	\dot{Q}_h = \frac{2}{h} \int dE \; T_{LR}(E) (E - \mu_h)  \left( f_L-f_R\right).
\end{equation}

As stated in the introduction, increasing the thermoelectric efficiency of nanostructures is pursued in two approaches. The first is to optimize the electronic properties in order to increase the generated electric power and the second is to reduce the thermal conductivity in order to decrease the heat exchange in the system.  
In this study, we have focused on the shape of the transmission function to maximize the generated electric power and efficiency. 
According to the thermodynamics laws, the maximum efficiency (Carnot efficiency) of heat engines operating at two different temperatures is equal to
\begin{equation}\label{eq:Carnot}
\eta_C=1-\frac{T_c}{T_h}
\end{equation} 
Carnot efficiency can only be achieved when the transmission function is equal to the delta function ($\delta(E-E_F)$) but such a transmission function leads to vanishing of the electric power output. 
The optimum efficiency at a given electric power is obtained for the case of boxcar shape transmission function ~\cite{Whitney14, Whitney15}. A boxcar transmission function acts as a band-pass filter and passes electrons in a special range of energy and blocks electrons out of this energy range. In the thermoelectric power generation process, the favorable electric current flows from hot lead to the cold side, and the current in the opposite direction is inappropriate. To enhance the electric power output and then thermoelectric efficiency, a boxcar transmission function only allowing ellectron flow in the desired direction while the destructive current in the opposite direction is eliminated.
The proper working temperature is determined by the width of the transmission window and should be in the order of $T\sim\frac{\Delta}{k_B}$, in which $T$ is the leads average temperature and $\Delta$ is width of the transmission window. The width of the transmission function determines the electrical output power and thermoelectric efficiency of the system. As $\Delta$ increases, the output power improves at the cost of decreasing efficiency and vise versa. Therefore, the optimal width depends on the required electrical output power and thermoelectric efficiency.

 \begin{figure}[]
  	\includegraphics[height=2.0\linewidth]{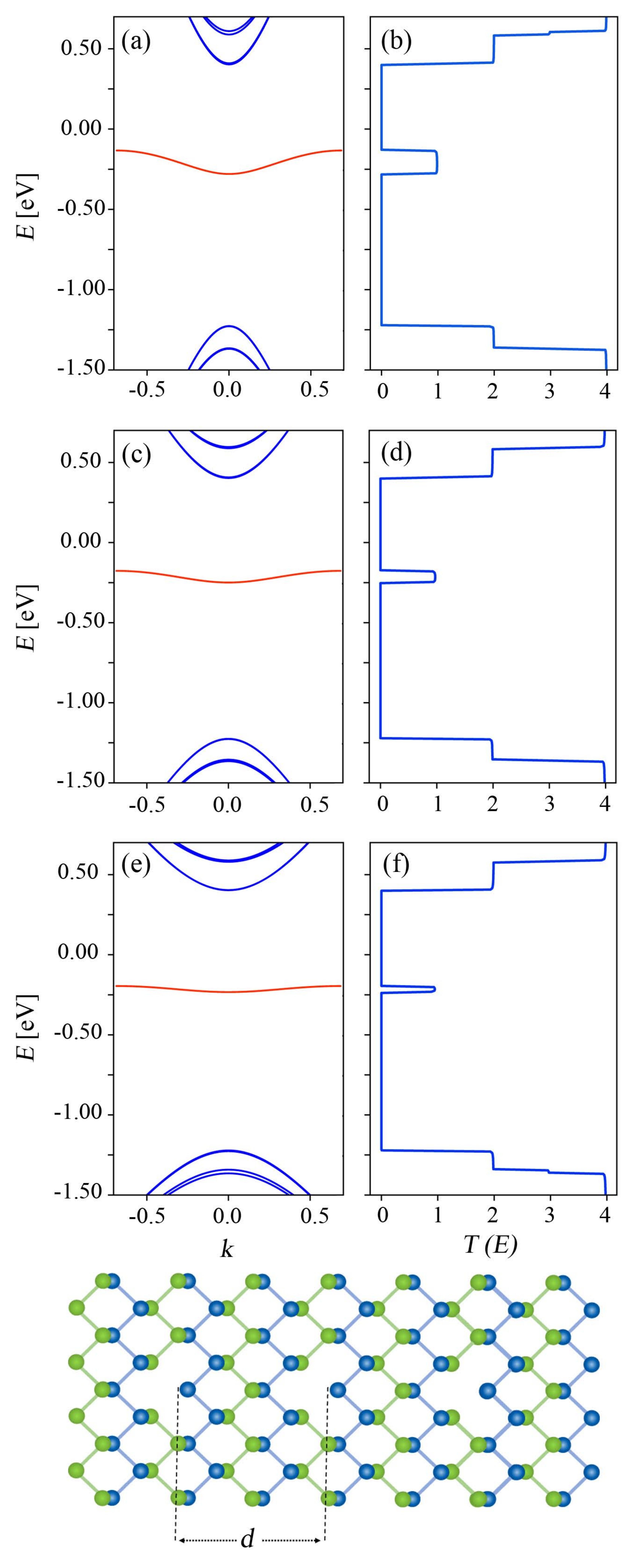}	
	\caption{\label{fig:T1atom}
	(a), (c), and (e) bandstructure close to the impurity band for cases where an atom is periodically removed. The distance between the removed atoms are $d=4b$, $6b$, and $8b$, respectively. The red lines indicate the impurity bands.
	(b), (d), and (f) transmission function close to the impurity bands corresponding to (a), (c), and (d).
	}
\end{figure}

   \begin{figure}[]
  	\includegraphics[height=2.0\linewidth]{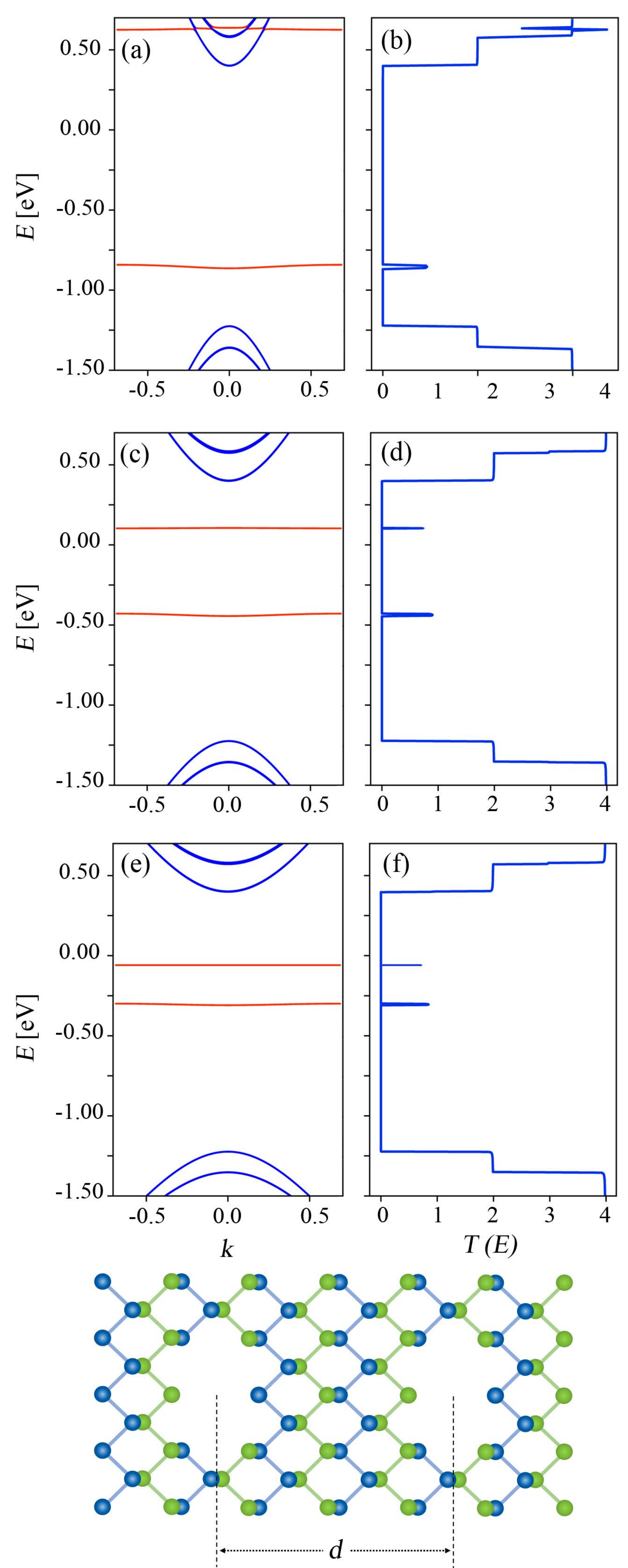}	
	\caption{\label{fig:T6atom}
	(a), (c), and (e) band structure close to the impurity bands of periodically created nanoholes of 6 atoms. The distance between the nanoholes are $d=4b$, $6b$, and $8b$, respectively. The red lines indicate the impurity bands.
	(b), (d), and (f) transmission function close to the impurity bands corresponding to (a), (c), and (d).
	}
\end{figure}

 \begin{figure*}[]
  	\includegraphics[height=0.95\linewidth]{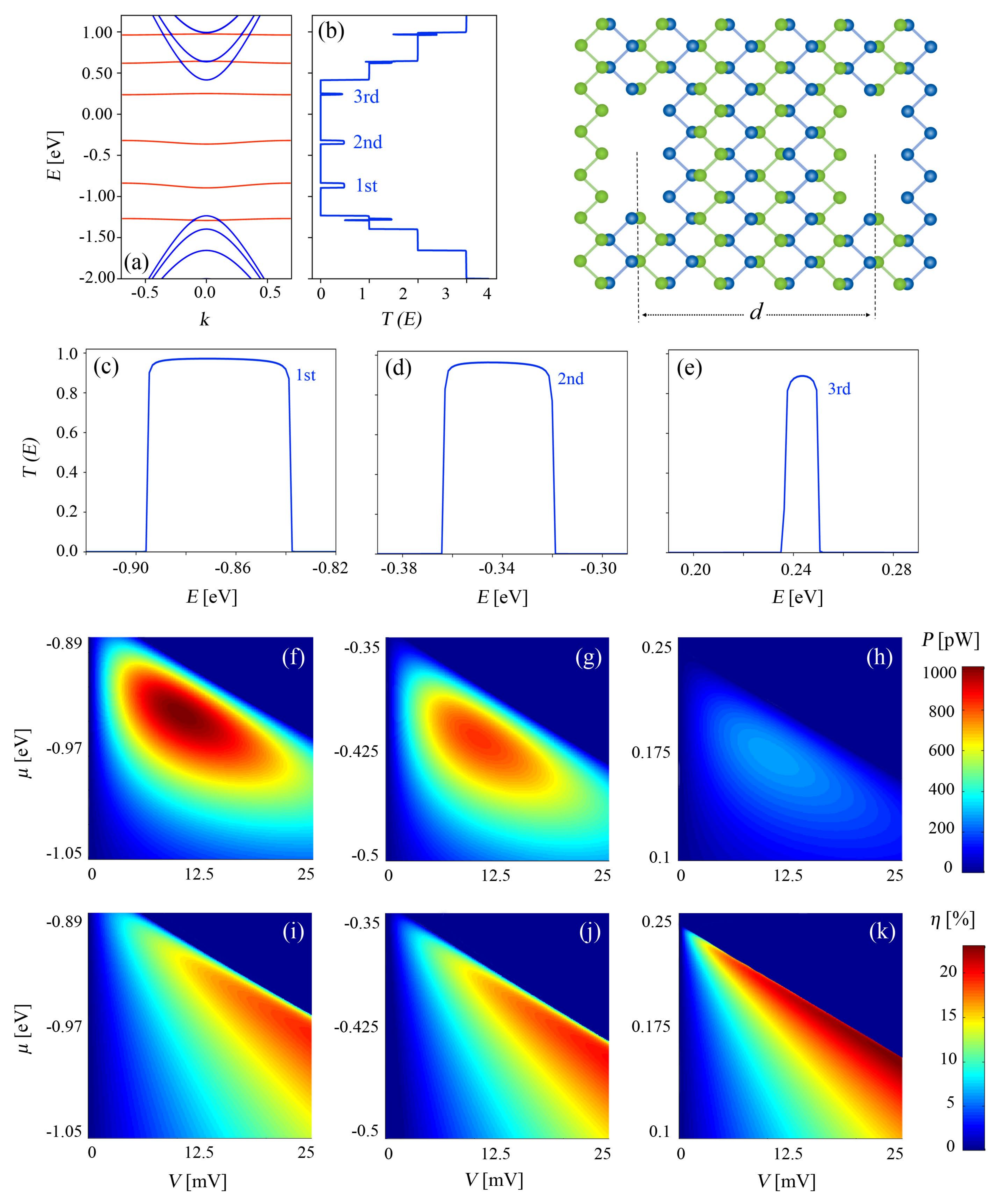}	
	\caption{\label{fig:T10atom}
	(a) band structure and (b) transmission function of APNR with periodically created nanoholes of 10 atoms. The distance between the nanoholes is $d=4b$. The red lines indicate the impurity bands. (c), (d), and (e) The transmission function close to each of the three bands that are labeled in part (b). (f), (g), and (h) electric power output and (i), (j), and (k) thermoelectric efficiency plotted on a color scale as a function of average chemical potential and bias voltage corresponding to the (c), (d), and (e), respectively. 
	}
\end{figure*}

\section{Results \label{sec:results}}
We assume that the temperature of the cold lead $T_c = 300~\mathrm{K}$ and for hot lead  $T_h = 400~\mathrm{K}$. So according to eq.~\ref{eq:Carnot}, the thermoelectric efficiency can never be higher than $25 \%$.

Figure ~\ref{fig:T1atom} (a) shows the energy bandstructure of an APNR with an array of periodic single atom vacancy defects. Here, the separation between defects is considered to be $d=4b$. 
The existence of an impurity band in the energy band gap, which is indicated by the red line, leads to the formation of a boxcar transmission window of width $\Delta$, Figure \ref{fig:T1atom} (b). 

Figures ~\ref{fig:T1atom}(c) and \ref{fig:T1atom}(e) show the bandstructures and figures~\ref{fig:T1atom}(d) and \ref{fig:T1atom}(f) illustrate the transmission functions corresponding to figures ~\ref{fig:T1atom}(a) and ~\ref{fig:T1atom}(b) when the spacing between vacancy defects is increased to $d=6b$ and $d=8b$, respectively. It is clear that as the distance between the vacancy defects increases, a narrower impurity transmission function is observed which is due to the reduction of overlap between the localized impurity states.

For larger nanoholes, a number of impurity bands and transmission windows may be observed, figure~\ref{fig:T6atom}. The number of impurity bands depends on the configuration of the vacant atoms in the nanoholes. The absence a single atom of type $A$ or $B$ (see fig.~\ref{fig:supercell}) results in an impurity band, while simultaneous removing of neighboring $A$ and $B$ atoms leads to no impurity band. 
The reason can be understood as follows: when there is only a single vacancy defect (on $A$ or $B$ sub-lattice), its associated wave-function can only exist on the other sub-lattice which has a zigzag boundary near the vacancy defect, but, when you have two nearby vacancies on two neighboring $A$ and $B$ sites, there is no zigzag edge on which the impurity state can exist~\cite{Amini2019_2,Amini2021}. 
Each nanohole in Figure~\ref{fig:T6atom} has 6 atoms, which consists of two pairs of neighboring $A$ and $B$ atoms and two single atoms, so two impurity bands are observed in the energy bandstructure. For the case of $d=4b$, only one of the impurity bands is within the energy bandgap and the other one, which overlaped with the bulk energy bands, does not operate as a proper transmisiion function for thermoelectric applications, figure \ref{fig:T6atom} (a). At larger distances between the nanoholes (figures~\ref{fig:T6atom} (c) and (e)), impurity bands move closer to each other and both of them are in the energy gap range. Similar to the case of monoatomic vacancy defects, for the cases with larger nanoholes, increasing the distance between the defects leads to narrowing of the transmission windows.

In Figure \ref{fig:T10atom}, we consider an APNR with 10-atomic vacancy defects and $d=4b$. In this case six impurity bands are formed and three of them are within the energy bandgap range, figure \ref{fig:T10atom} (a). The transmission windows corresponding to these impurity bands are labeled in Figure \ref{fig:T10atom} (b) and are ploted separately in figures \ref{fig:T10atom} (c), (d) and (e). Using each boxcar shape transmission window, we calculate electric output power $P$ and thermoelectric efficiency $\eta$. Figures~\ref{fig:T10atom} (f), (g) and (h) show P and figures~\ref{fig:T10atom} (i), (j) and (k) demonstrate $\eta$ as functions of the average chemical potential $\mu$ and bias voltage $V$, corresponding to figures~\ref{fig:T10atom}(c), (d) and (e), respectively. The maximum $P_{max}$ and $\eta_{max}$ can be achieved by accurately adjusting chemical potential $\mu$ and bias voltage $V$. One reason for choosing APNR with 10-atomic vacancy defects and $d=4b$ is that this case has three transmission windows with different widths in the energy bandgap range. In this case, by changing the chemical potential and the bias voltage, it is possible to choose one of the transmission windows that acts as a boxcar transmission function. The proper transmission window is selected based on the required $P$ and $\eta$. Another reason for choosing this case is the existence of more defects than smaller nanoholes, which further reduces the thermal conductivity of the system.

The widest transmission window has the highest maximum output power, $P^{1st}_{max} = 1012~pW$ and as expected, the maximum power decreases as the $\Delta$ decreases. here, $P^{2nd}_{max} = 831~pW$ and $P^{3rd}_{max} = 281~pW$. It is also clear that the maximum efficiency $\eta^{1st}_{max}$ decreases with increasing the width of transmission window. 
Since the output power corresponding to the maximum efficiency can be very small, the efficiency at maximum power in different cases is compared as a more appropriate performance metric than the maximum power and maximum efficiency. For the widest case, $\eta^{1st}_{Pmax}=12.5\%=0.5\eta_C$ and for the other transmission windows $\eta^{2nd}_{Pmax}=12.8\%$ and $\eta^{3rd}_{Pmax}=13.6\%$.
Although the $\eta_{Pmax}$ for the narrowest transmission window is slightly higher than the other cases, the widest transmission window is probably the best choice because of very large output power.

\section{Summery and conclusions \label{sec:conclusions}}
We have investigated the thermoelectric properties of an armchair phosphorene nanoribbon in the nonlinear response regime. We have shown that by creating an array of periodic vacancy defects, some impurity bands with boxcar shape transmission windows have been formed within the energy bandgap. Our results show that nanoholes size and their spacing have a major impact on the formation of transmission windows.  We have also calculated the electric output power and thermoelectric efficiency of an APNR with 10-atomic vacancy defects and have shown that such a structure can be used as a superb thermoelectric power generation module

\section{acknowledgment}
We would like to express our sincere thanks to the deputy of research and technology at the University of Isfahan and also to the Iran National Science Foundation (INSF) for their moral and financial supports. MA also acknowledges
useful discussions with L. Arrachea and Z. Nourbakhsh.

\appendix

\section{Proof of orthogonality property of wavefunctios defined in Eq.~(\ref{EQ7})}\label{APA}
For the readers convenience, we discuss the orthogonality property of wavefunctios defined in Eq.~(\ref{EQ7}) in this appendix.  
Without loss of generality, we consider $N=3$ bound states which can be described by $\mid \psi_0^{A}\rangle, \mid \psi_1^{A}\rangle, \mid \psi_2^{A}\rangle$ which do not fulfill the orthogonality condition,
\begin{equation}
 \langle\psi_{0}^A\mid\psi_{1}^A\rangle=\langle\psi_{1}^A\mid\psi_{2}^A\rangle=\langle\psi_{0}^A\mid\psi_{2}^A\rangle\neq0.
\end{equation}
We can now construct the basis functions $\mid\phi_{k_i}\rangle$ using the definition in Eq.~(\ref{EQ7}) as 
\begin{equation}
\begin{split}
 \mid\phi(k_0)\rangle &=\frac{1}{\beta_0}(e^{ik_0\times0}\mid\psi_{0}^A\rangle+e^{ik_0\times1}\mid\psi_{1}^A\rangle+e^{ik_0\times2}\mid\psi_{2}^A\rangle)\\
 \mid\phi(k_1)\rangle &=\frac{1}{\beta_1}(e^{ik_1\times0}\mid\psi_{0}^A\rangle+e^{ik_1\times1}\mid\psi_{1}^A\rangle+e^{ik_1\times2}\mid\psi_{2}^A\rangle)\\
 \mid\phi(k_2)\rangle &=\frac{1}{\beta_2}(e^{ik_2\times0}\mid\psi_{0}^A\rangle+e^{ik_2\times1}\mid\psi_{1}^A\rangle+e^{ik_2\times2}\mid\psi_{2}^A\rangle),
  \label{function3}
\end{split}
\end{equation}
and try to calculate their corresponding orthogonality explicitly.
Let us evaluate the first orthogonality condition
\begin{align}
 \langle\phi(k_0)\mid\phi(k_1)\rangle=&1/(\beta_0\beta_1)(\langle\psi_{0}^A\mid\psi_{0}^A\rangle+e^{2\pi i/3}\langle\psi_{0}^A\mid\psi_{1}^A\rangle\nonumber\\&+e^{4\pi i/3}\langle\psi_{\nu,0}^A\mid\psi_{\nu,2}^A\rangle+\langle\psi_{\nu,1}^A\mid\psi_{\nu,0}^A\rangle\nonumber\\&+e^{2\pi i/3}\langle\psi_{1}^A\mid\psi_{1}^A\rangle+e^{4\pi i/3}\langle\psi_{1}^A\mid\psi_{2}^A\rangle\nonumber\\&+\langle\psi_{2}^A\mid\psi_{0}^A\rangle)+e^{2\pi i/3}\langle\psi_{2}^A\mid\psi_{1}^A\rangle\nonumber\\&+e^{4\pi i/3}\langle\psi_{2}^A\mid\psi_{2}^A\rangle=0.
\end{align}
Applying the same procedure results in $\langle\phi(k_1)\mid\phi(k_2)\rangle=0$ and $\langle\phi(k_0)\mid\phi(k_2)\rangle=0$.
The same analysis can be applied for larger values of $N$.
This means that the wavefunctions of Eq.~(\ref{EQ7}) make an orthonormal basis.

\end{document}